\documentclass[a4paper,11pt]{article}
\pdfoutput=1 

\usepackage{jcappub} 

\usepackage[T1]{fontenc} 

\title{\boldmath New constraints on the  gamma-ray and  high energy neutrino fluxes from the   circumstellar interaction of SN 2023ixf}

\author[a]{Prantik Sarmah \note{Corresponding author.}}

\affiliation[a]{Department of Physics, Indian Institute of Technology Guwahati, Assam-781039, India}

\emailAdd{prantik@iitg.ac.in}

\abstract{The recent supernova, SN 2023ixf,  one of the closest observed  type II SNe has revealed the presence of a dense circumstellar material (CSM). Interaction of the SN ejecta with this dense CSM might create high energy protons of PeV energies through shock acceleration. These accelerated protons then colliding with the CSM (inelastic $pp$ collision)  can produce secondaries such as high energy gamma-rays and neutrinos. However, no gamma-rays and neutrinos have been detected by Fermi-LAT and IceCube from this event.  Fermi-LAT has placed an upper limit on the gamma-ray flux above $100$~MeV to be $2.6 \times 10^{-11}~\rm erg~cm^{-2}~s^{-1}$. On the other hand, IceCube's upper limit on muon neutrino flux is $7.3\times 10^{-2} ~\rm GeV~cm^{-2}$. {\color{black} Taking these limits into account and using the shock-CSM properties derived from multi-wavelength observations, we obtain new upper limits on the gamma-ray ($10^{-11}~\rm erg~cm^{-2}~s^{-1}$) and neutrino ($10^{-3}~\rm GeV~cm^{-2}$) fluxes from SN 2023ixf produced via the $pp$ interaction channel.} While we found the gamma-ray flux to be consistent with Fermi-LAT's upper limit, the neutrino flux is found to be about $2$ orders of magnitude smaller than the IceCube's upper limit.  We further analyse the detection prospects of such secondary signals from future SN 2023 like events with upcoming detectors, CTA and IceCube-Gen2 and found to have great discovery potential, if any similar event occurs within $7$~Mpc.}

\begin{document}
\maketitle
\flushbottom

\section{Introduction}
The  supernova, SN 2023ixf (distance, $6.9$~Mpc) in NGC 5457, classified as a type II SN is found to be more than $2$ magnitudes brighter than the typical type II SNe \citep{2023arXiv230604721J}. This bright emission is  produced by photo-ionization of a dense circumstellar material (CSM) around the progenitor star as  suggested by the flash spectroscopic analysis of  early emission from SN 2023ixf \citep{2023arXiv230610119B,2023arXiv230604721J,2023arXiv230703165H,2023arXiv230701268V}. Interaction of SN ejecta with this dense CSM may accelerate particles (e.g., protons) to PeV energies \citep{Schure:2013kya,Bell:2013kq,Cardillo:2015zda,Cristofari:2022kqv,Murase:2010cu,Petropoulou:2016zar}. These accelerated protons colliding with the CSM (inelastic $pp$ collision) can produce high energy secondaries such as gamma-rays and neutrinos \citep{Murase:2010cu,Murase:2013kda,Murase:2014bfa,Murase:2017pfe,Murase:2018okz,Petropoulou:2017ymv}. However, no gamma-rays and neutrinos have been detected by the present experiments, Fermi-Large Area Telescope (Fermi-LAT, gamma-ray detector) and IceCube (neutrino detector). Based on the first $14$ days of observation of SN 2023ixf, Fermi-LAT has suggested the gamma-ray flux above $100$~MeV to be not larger than $2.6 \times 10^{-11}~\rm erg~cm^{-2}~s^{-1}$  \citep{2023ATel16075....1M}. Similarly, IceCube has placed an upper limit of $7.3\times 10^{-2} ~\rm GeV~cm^{-2}$ on the muon neutrino flux in the energy range, $600$ GeV to $250$ TeV for a observation time of $+/- 2$~days  from the detection time of Zwicky Transient Facility \citep{2023ATel16043....1T}. In this work, we obtain the upper limits on the gamma-ray and neutrino fluxes from the $pp$ interaction channel  taking into account these experimental flux limits   and using properties of the shock-CSM model derived from light curve analysis, i,e., UV/optical observations~\citep{2023arXiv230610119B,2023arXiv230604721J}.

Even though there is fast growing evidence for dense CSMs around stars dying  as type II SNe  \citep{Smith:2014txa,2013MNRAS.430.1801M,Morozova:2016efp,Yaron:2017umb}, the production mechanism of such dense CSM is not well understood. Creation of such a dense CSM requires  large mass-loss rate, $\Dot{M}_{\rm W} \sim (10^{-3}-1)~\rm M_{\odot}~yr^{-1}$ occurring in the final stages of stellar explosion \citep{Smith:2014txa}. Production of these dense CSMs through usual stellar winds is debatable~\citep{Smith:2014txa}. Several alternate mass-loss mechanisms such as gravity waves, super-Eddington winds, binary interaction  have been proposed to explain the origin of such dense CSM \citep{Owocki_2004,10.1093/mnras/stx2251,10.1093/mnras/stw365,10.1093/mnras/stx1314,2021ApJ...906....3W,10.1093/mnras/staa255,2023arXiv230700727M}. 

Light curve modelling of SN 2023ixf suggests that  the dense CSM is produced by  $\Dot{M}_{\rm W} \sim 10^{-2}~\rm M_{\odot}~yr^{-1}$ with wind velocity, $v_{\rm W} \sim 50~\rm km~s^{-1}$ \citep{2023arXiv230604721J,2023arXiv230703165H,2023arXiv230610119B}.     This CSM is found to be confined within a radius of $10^{15}$~cm \citep{2023arXiv230604721J,2023arXiv230703165H,2023arXiv230615270S}. Beyond this radius, a sharp fall in the CSM density has been observed \citep{2023arXiv230604721J}. In addition, some studies of SN 2023ixf suggest a possible asymmetry in the CSM \cite{2023arXiv230607964S,2023arXiv230701268V}. The progenitor of SN 2023ixf is inferred to be a red supergiant (RSG) from observations \citep{2023arXiv230608678J,2023arXiv230514447P,2023arXiv230604721J}. The dense CSM is found to be composed of materials from RSG surface ejected via winds and/or violent outbursts \citep{2023arXiv230604721J,Smith:2014txa}. Although constituent elements of the dense CSM are not certain, observations suggest presence of hydrogen (H), helium (He), carbon (C) and nitrogen (N)\citep{2023arXiv230604721J}. Nevertheless, it is worthwhile to note that observations of a wide class of type II SNe have revealed their CSMs to be H-rich in nature \citep{Mauerhan:2018wes,2022arXiv221203313B,Milisavljevic:2015bli}.

Interaction of the SN ejecta with the dense CSM produces a strong shock expanding into the CSM (shock-CSM interaction) that may lead to  acceleration of particles to very high energies \citep{Murase:2010cu,Petropoulou:2016zar,2022JCAP...08..011S,2023arXiv230313576S,Schure:2013kya,Bell:2013kq,Cardillo:2015zda,Cristofari:2022kqv,Murase:2010cu,Petropoulou:2016zar}. The particle acceleration becomes efficient when the shock breaks out of the optically thick region of the CSM \citep{Petropoulou:2016zar}.  As mentioned above,   the accelerating particle  colliding with the CSM particles can create secondaries such as gamma-rays and neutrinos \citep{2022JCAP...08..011S}. The quest to detect these high-energy particles from SNe has become captivating because of the  high-energy neutrino background detected by IceCube~\cite{IceCube:2020wum} and the gamma-ray background observed by Fermi-LAT~\cite{Ackermann:2014usa}. Certainly, numerous analyses have been conducted on these high-energy particles originating from various types of supernovae~\cite{2022JCAP...08..011S,2023arXiv230313576S,Pitik:2021dyf,Pitik:2023vcg,Murase:2010cu,Murase:2013kda,Murase:2014bfa,Murase:2017pfe,Murase:2018okz,Petropoulou:2017ymv,Zirakashvili:2015tda,2018MNRAS.479.4470M,2014NuPhS.256...94M,2022MNRAS.511.3321C,2020MNRAS.494.2760C,2023ApJ...955L...9G}.

For an accurate estimation of these secondary fluxes, detail knowledge of the CSM composition is crucial. However, in this study, we only aim to obtain the upper limits of the gamma-ray and neutrino fluxes. Hence, we assume the CSM to be composed of protons only, with a wind density profile.  Using the model developed by \citep{Petropoulou:2016zar,Petropoulou:2017ymv,2022JCAP...08..011S}, we estimate the upper limits of the gamma-ray and neutrino fluxes produced by $pp$ interaction. The upper limits are found to be consistent with the  non-detection constraints from Fermi-LAT and IceCube. We further analyse the detection prospects of  such fluxes with upcoming detectors, Cherenkov Telescope Array (CTA) \citep{2011ExA....32..193A} and IceCube-Gen2 \citep{IceCube-Gen2:2020qha}, in the context of similar nearby SN events. These upcoming detectors are found to have great detection potential for SNe like SN 2023ixf, if any such SNe occurs within $7$~Mpc.

This paper is organised as follows. In Sec.~\ref{sec:model}, we  describe the model of shock-CSM interaction, shock acceleration of protons and production of secondary gamma-rays and neutrinos. We also discuss the procedure to compute these secondary fluxes. In Sec.~\ref{sec:fluxes}, we compute the secondary fluxes and discuss their non-detection in Fermi-LAT and IceCube. We present our upper limits on the secondary fluxes with multi-messenger implications and discuss future detection prospects for similar events in Sec.~\ref{sec:results}. In Sec.~\ref{sec:sec_radiation}, we provide an estimate of the multi-frequency radiation from secondary electrons. Finally, we conclude the paper in Sec.~\ref{sec:conclusion}.

\section{Modeling gamma-ray and neutrino emission in CSM interaction}
\label{sec:model}
Gamma-rays and high energy neutrinos can be produced through the decay of neutral pion ($\pi^{0}\longrightarrow 2 \gamma$) and  charged pions ($\pi^{\pm}\longrightarrow \mu^{\pm}+\nu_{\mu}$ and $\mu^{\pm}\longrightarrow e^{\pm} + \nu_{\mu} +\nu_{e}$) created in the interaction of
 shock accelerated protons with low energy CSM protons ($pp$ collisions) \citep{Petropoulou:2016zar,2022JCAP...08..011S,Kelner:2006tc}. Note that, we treat neutrinos and anti-neutrinos as same here. The fluxes of these secondary particles depend on the density profile of the CSM and the  spectrum of the shock accelerated protons. In the following, we describe the modelling of these ingredients and the method to compute the fluxes of the secondaries. 

We assume a spherically symmetric CSM described by the wind density profile as a function of radius $r$, 
\begin{equation}
     n_{\rm CSM}(r) = \frac{\Dot{M}_{\rm W}}{4 \pi v_{\rm W} m_{\rm H} r^2}~,
\end{equation}

 where $\Dot{M}_{\rm W}$ and $v_{\rm W}$ are mass-loss rate and wind velocity, respectively.   The CSM is assumed to be confined between the inner radius $r_{\rm i}$ and the outer radius $r_{\rm o}$, and composed of hydrogen ( mass  $m_{\rm H}$) only.   SN ejecta created by SN explosion interacting with CSM produces a fast forward shock expanding into the CSM which accelerates protons to high energies  \citep{Petropoulou:2016zar}. The  shock speed, $v_{\rm sh}$ can be assumed to be constant as it varies slowly with the shock radius \citep{Petropoulou:2016zar}.    For the SN 2023ixf, $v_{\rm sh}$ is found to be larger than  $8.5 \times 10^3~\rm km\ s^{-1}$ \citep{2023arXiv230604721J}. 

The shock accelerated protons with energy, $E_{\rm  p}$  can be described by a power law distribution with an exponential cut-off of the form, 

\begin{equation}
    \mathcal{Q}_{\rm p}(E_{\rm p},r) \propto E_{\rm p}^{-2} \exp \left(\frac{-E_{\rm p}}{E_{\rm p,max}(r)} \right)~,
    \label{eq:proton}
\end{equation}
where  $E_{\rm p,max}(r)$ is  the maximum proton energy \citep{Petropoulou:2016zar,Murase:2010cu,Murase:2017pfe}. The maximum proton energy $E_{\rm p,max}(r)$ governs the spectral shape of this distribution at higher energies. $E_{\rm p,max}(r)$ is determined by the acceleration time scale  ($t_{\rm acc}(r)$) competing with different energy loss timescales,  mainly $pp$ collision loss ($t_{\rm pp}(r)$) and dynamical or adiabatic loss ($t_{\rm ad}(r)$), i.e., $t_{\rm acc}(r)=\min[t_{\rm pp}(r),t_{\rm ad}(r)]$  \citep{Petropoulou:2016zar}. Other loss processes such as photo-pion production, synchrotron, inverse Compton are negligible \citep[][]{2022JCAP...08..011S}. The acceleration time scale  
is defined as $t_{\rm acc}(r)= 6 E_{\rm p} c/ e B v_{\rm sh}^2$ {\color{black} in the Bohm limit \footnote{It is worthwhile to note that the assumption of Bohm diffusion is applicable to  spectrum of accelerated
particles with spectral index $\sim 2$ and requires the Bell modes to be excited in order to distribute the same power across all wave modes ~\cite{10.1111/j.1365-2966.2004.08097.x,2009MNRAS.392.1591A}. }}, where $B$ is the magnetic field strength of the post shock CSM. {\color{black} Particle acceleration in dense CSM can be efficient due to strong magnetic field}.   Magnetic field is expected to be amplified due to instabilities (e.g., Bell instability~\cite{10.1111/j.1365-2966.2004.08097.x}) caused by upstream particles~\cite{10.1111/j.1365-2966.2004.08097.x,10.1093/mnras/stt179,Schure:2013kya,Schure:2012du}. In addition, small scale strong magnetic field turbulence can also be possible in such environment due to phenomena such as plasma instabilities and turbulent dynamo effect~\cite[see e.g.,][]{2009ApJ...695..825I,2012ApJ...744...71I}. As these processes are uncertain, we determine the magnetic field by assuming the magnetic energy density ($B^2/(8\pi *)$) to be a fraction ($\epsilon_{\rm B}$) of the post shock kinetic energy~\cite{Murase:2013kda,Petropoulou:2016zar}. Thus, the magnetic field is  given by $B=3/2 [4 \pi \epsilon_{\rm B} m_{\rm p} n_{\rm CSM}(r) v_{\rm sh}^2]^{1/2}$. This shows that the magnetic field is a function of $r$. The uncertainties in the CSM density and shock velocity can lead to uncertainties in the magnetic field. In addition, the magnetic field can also vary due to radial evolution of different instabilities/turbulence. This can also be a possible source of uncertainty in the magnetic field.   We assume the uncertainty in the magnetic field to be due to $\epsilon_{\rm B}$ only and therefore, we treat $\epsilon_{\rm B}$ as a free parameter. 
Typically, for Type II SNe with dense CSM, $\epsilon_{\rm B}$ is found to be in the range $10^{-3}-10^{-1}$ \citep{2023arXiv230609311B,2014ApJ...781...42O,2009ApJ...690.1839C,2013ApJ...768...47O,2000ApJ...536..195C}. Note that large value of $\epsilon_{\rm B} \sim 0.1$  implies strong magnetic field pressure in the upstream shock. If the pressure due to magnetic field becomes comparable to the thermal pressure, it may result in a steeper spectrum of accelerated particles due to dynamical effects, i.e, spectral index larger than $2$~\cite{Caprioli:2008aq}. However, we choose to keep the spectral index to be $2$ based on the standard diffusive shock acceleration theory as our goal is to obtain maximum secondary (gamma-rays and neutrinos) fluxes.

The $pp$ collision time scale is defined as $t_{\rm pp}(r)= [\kappa_{\rm pp} \sigma_{\rm pp} n_{\rm CSM}(r) c]^{-1}$, where $\kappa_{\rm pp} =0.5$ is the proton inelasticity and $\sigma_{\rm pp}$ is the $pp$ interaction cross-section \citep{Kelner:2006tc,2022JCAP...08..011S}. The adiabatic loss timescale  is given by $t_{\rm ad} (r) \sim r/{v_{\rm sh}}$. Thus, $E_{\rm p,max}(r)$ as a whole depends on the CSM density ($n_{\rm CSM}(r)$), the shock velocity ($v_{\rm sh}$) and the magnetic field parameter ($\epsilon_{\rm B}$). Note that larger  $v_{\rm sh}$ and $\epsilon_{\rm B}$ result in larger $E_{\rm p,max}(r)$ whereas larger $n_{\rm CSM}(r)$ yields smaller $E_{\rm p,max}(r)$.  
The normalization of the primary proton spectra   $\mathcal{Q}_{\rm p}(E_{\rm p},r)$ is obtained by the total energy of the accelerated protons which is considered to be a fraction, $\epsilon_{\rm p}$ of the shock kinetic energy. The shock kinetic energy is given by $E_{\rm s}= (9 \pi/8) m_{\rm p} v_{\rm sh}^2 r^2 n_{\rm CSM}(r)$~\citep[][]{Petropoulou:2016zar}.  The parameter $\epsilon_{\rm p}$ is found to be in the range $10^{-2}-10^{-1}$ \citep{2014ApJ...781...42O,2009ApJ...690.1839C,2013ApJ...768...47O,2000ApJ...536..195C} and hence treated as a free parameter.  

The proton spectra, $\mathcal{Q}_{\rm p}(E_{\rm p},r)$ evolves with time  due to the radial evolution of the CSM density and the different processes discussed above. { \color{black} The time ($t$) evolution of the spectra is obtained by assuming  the linear  relation, $r=v_{\rm sh} t$. The evolution of the shock is in general non linear and this relation holds because the shock velocity changes slowly with time~\cite[see e.g.,][]{Petropoulou:2016zar,Murase:2017pfe}.} The evolution of the proton distribution denoted by $\mathcal{N}_{\rm p}(E_{\rm p},r)$ can be obtained by the following equation \citep{Sturner_1997,Petropoulou:2016zar},

\begin{align}
\frac{\partial \mathcal{N}_{\rm p}(E_{\rm p},r)}{\partial r}+\frac{\mathcal{N}_{\rm p}(E_{\rm p},r)}{v_{\rm sh} t_{\rm pp}(r)}-\frac{\partial}{\partial E_{\rm p}}\left[\frac{E_{\rm p} \mathcal{N}_{\rm p}(E_{\rm p},r)}{r}\right] = \mathcal{Q}_{\rm p}(E_{\rm p},r)\ .
\label{proton_de}
\end{align}

The second and third term in the left hand side corresponds to $pp$ collision loss and the adiabatic loss respectively. {\color{black} These two loss terms govern the evolution of the proton distribution. The adiabatic loss term corresponds to the protons escaping due to adiabatic expansion of  the shocked shell of the CSM~\cite{Petropoulou:2016zar}. The $pp$ collision loss term can be assumed to be a catastrophic energy loss process as a single $pp$ collision causes  significant energy loss ($\kappa_{\rm pp}=0.5$) to the accelerated protons~\cite{1997ApJ...490..619S}. This ensures that the $pp$ loss term  behaves as an escape term in the above evolution equation. Therefore, the evolution of the accelerated proton distribution becomes just a number changing process under such condition~\cite{1995A&A...295..613M}. } This  proton distribution ($\mathcal{N}_{\rm p}(E_{\rm p},r)$) can now be used to compute the secondary spectra i.e, gamma-rays and   neutrinos, $\mathcal{Q_{\rm i}}(E_{\rm i},r)$, where $E_{\rm i}$ is the   energy of the secondaries with $i=\gamma,~\nu_{e},~\nu_{\mu}$. $\mathcal{Q_{\rm i}}(E_{\rm i},r)$ depends on the $pp$ collision cross-section~\cite{Kelner:2006tc}, the production rate of the secondaries~\cite{Kelner:2006tc}, the CSM density and the shock velocity \citep{2022JCAP...08..011S}.  These secondaries are produced in the CSM at different radii  and therefore, they escape the CSM at different radii which  corresponds to  different time. The escape timescale of these secondaries is defined as $t_{\rm esc} (r) =r/ (4c)$.  {\color{black} The factor $4$ is added   to incorporate compression of the downstream CSM due to strong forward shock~\cite{Sturner_1997,Petropoulou:2016zar}.} Thus, the secondary spectral distribution, $ N^{\rm S}_{\rm i}(E_{\rm i},r)$ at source (S) can be obtained by \citep{Petropoulou:2017ymv},

\begin{equation}
    \frac{\mathrm{d} N^{\rm S}_{\rm i}(E_{\rm i},r) }{\mathrm{d}r} + \frac{N^{\rm S}_{\rm i}(E_{\rm i},r) }{v_{\rm sh} t_{\rm esc}(r)} = \mathcal{Q}_{\rm i}(E_{\rm i},r)\ .
    \label{eq:Secondary_de}
\end{equation}

Neutrinos being weakly interacting can escape the CSM without any losses. However, the gamma-rays will attenuate due to pair production on low energy  SN photons.  The attenuation due to this loss can be estimated by the factor, $\exp{\left( -\tau_{\gamma\gamma}\right (E_{\gamma},r))}$, where $\tau_{\gamma\gamma}(E_{\gamma},r)$ is the optical depth of the SN photons \citep{2022JCAP...08..011S}. The optical depth depends on the density profile of SN photons and the size of the SN photon environment.  The SN photon density can be estimated with two ingredients,  the observed  SN peak luminosity ($L_{\rm SN,pk}$) and the average energy ($\epsilon_{\rm av}$) \citep{2022JCAP...08..011S}. The radial variation of the photon density is assumed to be of the form, $r^{-2}$. This radial density profile results in radial decline of $\tau_{\gamma\gamma}(E_{\gamma},r)$. The peak luminosity of SN 2023ixf is found to be in the range, $10^{42}-10^{43}$~$\rm egr/s$ and the average energy is $\mathcal{O}(1)$~eV \citep{2023arXiv230604721J}. Note that the variation of the optical depth with radius will result in variation of the  attenuation. Since the optical depth decreases with increasing radii, the attenuation will be smaller at larger radii.  Apart from pair production loss, gamma-rays might suffer from losses due to interaction with ambient nuclei. However, the attenuation due to this process is found to be negligible \citep{2022JCAP...08..011S}.

Now, with these information one can estimate the expected  secondary (gamma-ray and neutrino) fluxes at Earth. The secondary fluxes at Earth from a source at distance, $d_{\rm L}$ will  be modified by the propagation factor $1/4 \pi d_{\rm L}^2$.   Thus, the secondary gamma-ray and neutrino   fluxes at Earth are given by,

\begin{equation}
    \phi_{\rm i}(E_{\rm i},r)=   \frac{  N^{\rm S}_{\rm i}(E_{\rm i}^{\prime},r) }{4 \pi d_{\rm L}^2(1+z)^2 t_{\rm esc}(r)}\ ,
\label{eq:NuFlux}
\end{equation}

where $E_{\rm i}^{\prime}= (1+z) E_{\rm i}$ and $z$ is the cosmological redshift. For the source of our interest which is located at $6.9$~Mpc, the effect of  $z$ on the flux is negligible. 

These secondaries can suffer from losses during propagation to Earth due to interaction with different intervening backgrounds like the Cosmic Microwave Background (CMB) and the Extra-galactic Background Light (EBL).
The propagation losses of gamma-rays on these backgrounds   is found to negligible for small distances, i.e., $\mathcal{O}(10)$~Mpc \citep{2022JCAP...08..011S}. While the neutrinos do not interact with these backgrounds, but can undergo flavour conversion during propagation. The flavour conversion effect is given by the flavour ratio $\nu_{e} : \nu_{\mu} : \nu_{\tau} = 1 : 1 :1$ \citep{PhysRevD.90.023010}. To obtain the neutrino flux of any specific flavour, one needs to consider one third of the all flavour flux.


 \begin{table}
    \centering
    \begin{tabular}{|c|c|c|c|}
    \hline
    {\bf Parameters} & {\bf Typical value}\\
    \hline
    \hline
    $\Dot{M}_{W}$ ($\rm M_{\odot}~yr^{-1}$)  & $10^{-2}$\\
    \hline
    $v_{\rm W}$ $~(\rm km\ s^{-1})$ & $50$\\
    \hline 
    $r_{\rm i}~(\rm cm)$ & $1\times 10^{14}$ \\
    \hline 
    $r_{\rm o}~(\rm cm)$ & $ 1 \times 10^{15}$ \\
    \hline
    $v_{\rm sh}~(\rm km\ s^{-1})$ & $10^4$ \\  
    \hline
    $\epsilon_{\rm B}$ & $ 10^{-1}$ \\
    \hline
    $\epsilon_{\rm p}$ & $  10^{-1}$ \\
    \hline
    $L_{\rm SN,pk}$ ($\rm erg s^{-1}$) & $10^{42}$ \\
    \hline
    $\epsilon_{\rm av}$ eV & $1$\\
    \hline
    $d_{\rm L}$ & $6.9$~Mpc \\
    \hline
    Declination & $\sim 54^{\circ}$\\
    \hline
    \end{tabular}
    \caption{List of the characteristic model  parameters. The parameters are optimised to obtain maximum gamma-ray and neutrino fluxes.}
    \label{tab:parameters}
\end{table}

\begin{figure*}
\includegraphics[width=0.49\textwidth]{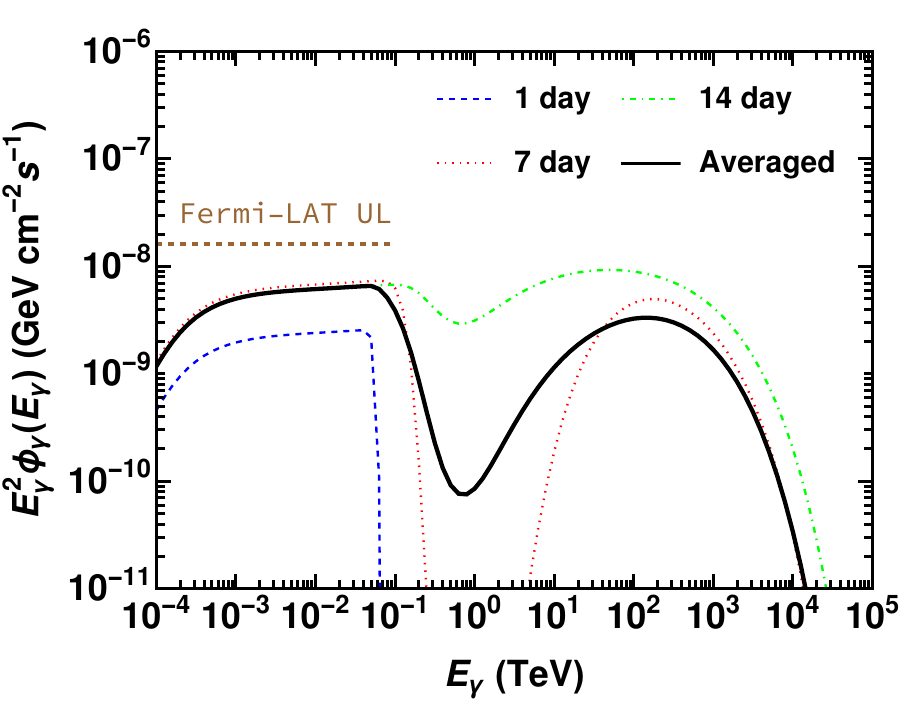}
\includegraphics[width=0.49\textwidth]{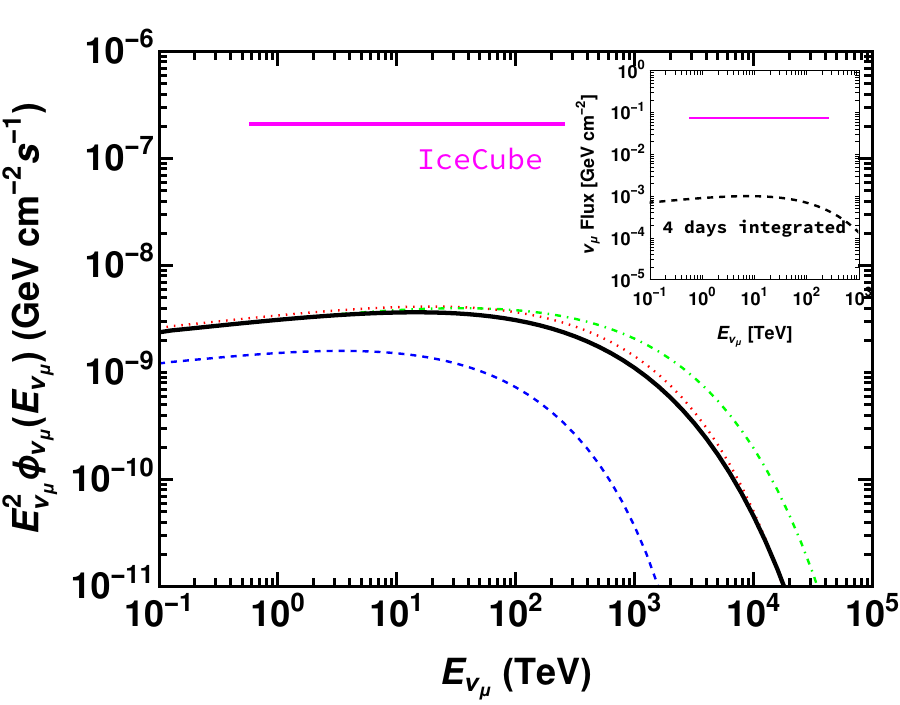}
\caption{Gamma-ray (left) and neutrino (right) fluxes at different days of shock-CSM interaction. The black curves show the fluxes averaged  over $14$ days. The brown dashed line in the left panel shows the Fermi-LAT upper limit, whereas the magenta line in the right panel shows the IceCube's upper limit on the muon neutrino flux for $4$ days of observation. The inset plot in the right panel shows the time integrated flux for 4 days (black dashed) with the IceCube's upper limit. The fluxes are consistent with the non-detection results reported by these experiments.}
\label{Fig:Fluxes}
\end{figure*}

\section{Estimation of gamma-ray and neutrino fluxes} \label{sec:fluxes}
Using the model described above, we now compute the fluxes of gamma-rays and neutrinos from the CSM interaction of SN 2023ixf. The fluxes depend on the different parameters characterising the shock and the CSM models. We select a set of   parameter values that are based on analyses of the observed  data~\citep{2023arXiv230604721J,2023arXiv230703165H,2023arXiv230610119B}. These parameter values are listed in Table~\ref{tab:parameters}. The parameters are optimised in such a way that they produce the largest fluxes of gamma-rays and neutrinos  as our goal is to obtain the upper limits on the fluxes. {\color{black}  The large value of  $\epsilon_{\rm B}$ in Table~\ref{tab:parameters}  taken from~\citep{2023arXiv230609311B}  is comparatively larger than general expectation for interacting SNe~\citep{2014ApJ...781...42O,2009ApJ...690.1839C,2013ApJ...768...47O,2000ApJ...536..195C}. We allow this large value as we want to obtain the maximal secondary fluxes. Lower values  of $\epsilon_{\rm B}$ will yield lower value of  maximum proton energy. This will in turn result in  smaller cut-off energies for the secondary fluxes.  }

The fluxes of  gamma-rays and neutrinos  evolve with radii according to Eq.~\ref{eq:Secondary_de}. The fluxes for different radii correspond to different time of shock-CSM interaction. Understanding this temporal evolution is crucial for probing these fluxes. Therefore, to illustrate this effect we compute the secondary fluxes at three different time of CSM interaction, i.e, 1 day, 7 days and 14 days. The maximum time is chosen to be 14 days based on the 14 days of observation time of Fermi-LAT. This also corresponds to the outer radius of CSM.  The predicted upper limit on the gamma-ray flux by the Fermi-LAT collaboration  is about $1.6\times 10^{-8}$ $\rm GeV~cm^{-2}~s^{-1}$ \citep{2023ATel16075....1M}. On the other hand, the upper limit  on the muon neutrino flux by IceCube in the energy range $600$ GeV to $250$ TeV is  $7.3\times 10^{-2} ~\rm GeV~cm^{-2}$ (equivalent to $2.1 \times 10^{-7} ~\rm GeV~cm^{-2}~s^{-1}$) for a search of $+/-2$ days~\citep{2023ATel16043....1T}.  
{\color{black}Using the constraints on the shock-CSM properties obtained from observed data \citep{2023arXiv230610119B,2023arXiv230604721J,2023arXiv230703165H,2023arXiv230701268V}, we  estimate the gamma-ray and neutrino fluxes produced by the shock-CSM model. While doing this, we make sure that the secondary fluxes corresponding to  our parameter choices do not exceed the limits from Fermi-LAT and IceCube.}

The left panel of Fig.~\ref{Fig:Fluxes} shows the gamma-ray fluxes at different days, i.e, $1$ (blue dashed), $7$ (red dotted) and $14$ (green dot-dashed). The flux corresponding to $1$ day being smaller than the remaining fluxes represents the beginning of the shock-CSM interaction. The flux grows  with time which is depicted by the fluxes at $7$ and $14$ days respectively. This is due to the radial decline of the CSM. As discussed above, the gamma-rays suffer from attenuation due to pair production on SN photons. The large dips in the fluxes show this attenuation effect. The sizes of the dips are different at different days as the optical depth of the SN photons varies with radii. On the other hand, all the fluxes  fall rapidly at higher energies. This is because the maximum proton energy, $E_{\rm p,max}(r)$ gives rise to a high energy cut-off to the secondary fluxes.  The variation of  $E_{\rm p,max}(r)$ with $r$ resulted in different cut-off energies at different time for the secondary gamma-rays. As the gamma-ray fluxes are different at different times of shock-CSM interaction, we also compute the average flux over the duration of 14 days.  The average flux is shown by the thick black curve. In addition to these fluxes, we depict the Fermi-LAT upper limit ($< 1.6 \times 10^{-8}~\rm GeV~cm^{-2} ~s^{-1}$) on the gamma-ray flux above $100$~MeV by the brown dashed line.   This flux limit is plotted up to the maximum detectable energy of Fermi-LAT, i.e, $\sim 300$~GeV. The predicted gamma-ray fluxes with the shock-CSM model is found to be well below this upper limit.

Concurrently, we also show the muon neutrino fluxes at different days, $1$ (blue dashed), $7$ (red dotted) and $14$ (green dot-dashed) in the right panel of Fig.~\ref{Fig:Fluxes}. Similar to the gamma-rays, the neutrino flux at $1$ day corresponds to the beginning of shock-CSM interaction. The neutrino flux then grows with time as shown by the fluxes ar $7$ and $14$ days. The variation of $E_{\rm p,max}(r)$ with $r$ also resulted in different cut-off energies for the neutrino fluxes. Like the gamma-ray case, the average muon neutrino flux is shown by the thick black curve. In addition to the muon neutrino fluxes, we also plot the IceCube's upper limit on the muon neutrino flux by the magenta line. Our predicted neutrino fluxes are about $2$ orders of magnitude smaller than this limit. For consistency, we also depict the time integrated muon neutrino flux for $4$ days (which is the observation time of IceCube~\citep{2023ATel16043....1T}) together with the IceCube's upper limit in the inset plot of the right panel of Fig.~\ref{Fig:Fluxes}.  Thus, the neutrino flux prediction is consistent with the non-detection of neutrinos by IceCube from SN 2023ixf.

\begin{figure}
    \centering
    \includegraphics[width=0.7\textwidth]{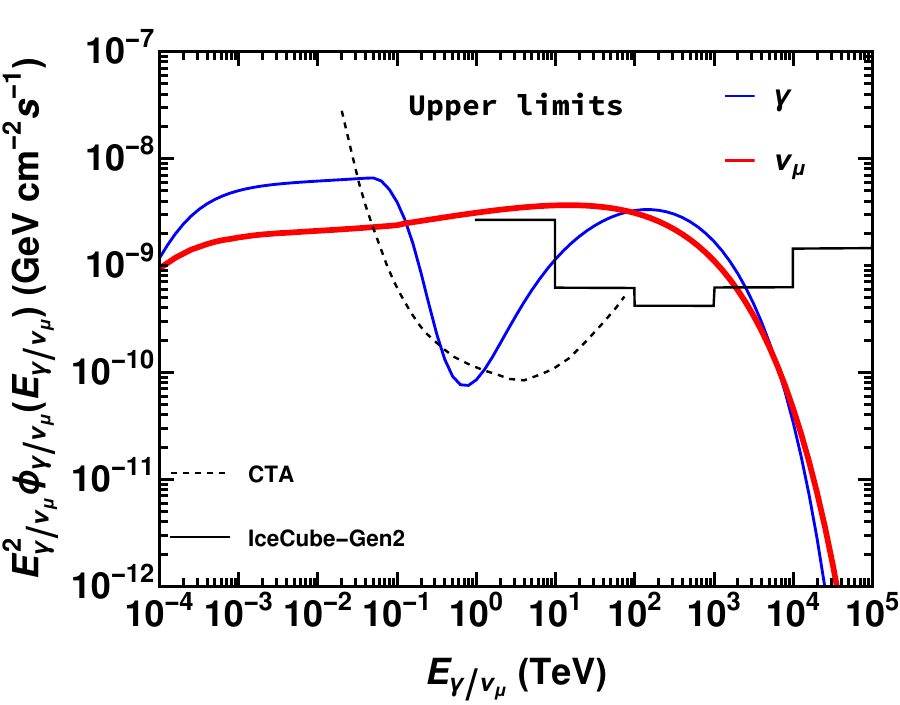}
    \caption{Upper limits of the gamma-ray (blue thin) and neutrino (red thick) fluxes together with sensitivity of CTA (black dashed) and IceCube-Gen2 (black continuous). This shows that these detectors will be able to probe similar nearby events in future.}
    \label{fig:final}
\end{figure}

\section{Multi-messenger constraints and Future perspectives} \label{sec:results}

{\color{black} 
The estimated secondary gamma-ray and neutrino fluxes  can provide crucial multi-messenger constraints on the properties of the shock-CSM interaction.
For any firm constraints on the model parameters, detection of these fluxes across different energies is crucial. For instance, the detection of the fluxes at the extreme high energies will allow us to probe the maximum proton energy, $E_{\rm p,max}(r)$. This will in turn  help us to understand the underlying acceleration mechanism constraining parameters such as $v_{\rm sh}$ and $\epsilon_{\rm B}$.  On the other hand, the detection at GeV-TeV range will constrain the CSM properties such as density. It may also provide constraints on the parameter $\epsilon_{\rm p}$ as the fluxes are proportional to  $\epsilon_{\rm p}$.

We now provide a multi-messenger analysis of the secondary fluxes using the experimental limits (Fermi-LAT and IceCube) .
For this, we compare the estimated secondary fluxes with the experimental limits  as shown in Fig.~\ref{Fig:Fluxes}. The gamma-ray flux being comparable to the Fermi-LAT  limit turns out to be useful for constraining the different model parameters.  However as the neutrino flux being much lower than the IceCube's limit, the corresponding constraints are found to be non-conclusive. Nevertheless, we have taken into account these constraints while choosing the values of the parameters given in Table~\ref{tab:parameters}. Note that these choices are not unique, one may find a different set parameter values producing similar secondary fluxes. Therefore, detection of the secondary fluxes is essential  in order to firmly constrain these parameters.

The time-averaged fluxes of gamma-rays and neutrinos plotted in Fig.~\ref{Fig:Fluxes} being closer (but smaller) to the maximum fluxes (fluxes at 7 and 14 days of interaction) can be useful for  obtaining the maximal detection potential. We adopt these time-averaged fluxes as conservative upper limits on the gamma-ray and neutrino fluxes.  Hence, this indicates that the gamma-ray and high energy neutrino fluxes from SN 2023ixf are not larger than $7\times 10^{-9}~\rm GeV~cm^{-2}~s^{-1}$ and $4\times 10^{-9}~\rm GeV~cm^{-2}~s^{-1}$, respectively if they are produced by $pp$ interaction. Note that,  these multi-messenger constraints  are weak at present as these fluxes were not detected by Fermi-LAT and IceCube. 
However, highly sensitive upcoming detectors such as CTA and IceCube-Gen2 will be able to detect the fluxes of gamma-rays and neutrinos from similar future events and will put stringent constraints on the model parameters. Therefore, we analyse the detection potential of these future detectors in the following.


We have depicted the future detection potential of SN 2023ixf like events in 
Fig.~\ref{fig:final}. 
In order to demonstrate this, we have plotted plotted the upper limits of the gamma-ray and high energy neutrino fluxes from SN 2023ixf together the sensitivities of CTA (black dashed, $50$ hour)~\citep{2011ExA....32..193A} and IceCube-Gen2 (black continuous, $10$ year \footnote{Note that we only show the available IceCube-Gen2 sensitivity to demonstrate future possibilities. For accurate prediction, one needs to compute the number of events subtracting the background effects which is beyond the scope of this work.})~\citep{IceCube-Gen2:2020qha}.  This  infers  for  a future event within $7$~Mpc, these detectors will be  able to detect  the gamma-ray and neutrino fluxes if it occurs during the run time of these detectors. Such detection in the future will help in understanding particle acceleration in early SN shock as well as properties of the CSM. In addition, the CTA is sensitive to the dips in the gamma-ray fluxes  in the energy range  $(10^1-10^5)$~GeV. Thus, CTA will also be able to probe gamma-ray attenuation at source.
}

{\color{black} The gamma-ray and neutrino fluxes presented in this work are computed for a primary proton power law index $=2$ (see Eq.~\ref{eq:proton}). However, steeper proton spectra  (i.e, power law index larger than $2$) are possible due to  effects such as  magnetic feedback, finite velocity of scattering centres in the acceleration environment~\cite{Caprioli:2008aq,Cristofari:2022kqv}. In fact, there are numerous indirect evidences that suggest steeper spectra for accelerated particles~\cite{Chevalier:2006vy,2010ApJ...725..922S,2016ApJ...818..111K,2017MNRAS.472.2956A}.  For steeper proton spectra, the spectral shape of the secondary gamma-ray and neutrino fluxes will also change.  Considering the Fermi-LAT upper limit  and the benchmark parameter values in Table~\ref{tab:parameters}, one can accommodate larger power law index  up to  $\sim 2.3$. For this specific scenario, the gamma-ray flux will be larger (compared to power law index $=2$) in the Fermi-LAT energy range and the upper limit on the gamma-ray flux from $pp$ channel will be the same as the Fermi-LAT upper limit.  Consequently, the neutrino flux will become steeper at higher energies (smaller flux compared to power law index $=2$) and will have weak sensitivity in IceCube-Gen2. Nevertheless, the gamma-ray flux will  be still detectable in CTA. 

}

\section{Multi-frequency radiation from secondary electrons}
\label{sec:sec_radiation}
{\color{black}
In the above analysis, we have demonstrated the gamma-ray flux from the decay of  $\pi^0$  produced in $pp$ interaction. This $pp$ interaction can also produce secondary electrons (as well as positrons but we do not intend to distinguish particle and anti-particle) via the decay of charged pions, $\pi^{\pm}$. These secondary electrons can produce non-thermal multi-frequency (ultra-violet, X-rays, and gamma-rays) photon flux  via synchrotron radiation.  In this section, we briefly describe this process and provide an estimate of the multi-frequency photon flux for our benchmark parameter values listed in Table~\ref{tab:parameters}.

The intensity, $I_{\nu}$ (in units of $\rm ergs~cm^{-2}~s^{-1}~Hz^{-1}$) of the synchrotron spectrum can be obtained by the following formula~\cite{Petropoulou:2016zar},

\begin{equation}
    I_{\nu} = \frac{j_{\nu}}{\alpha_{\nu}^{\rm sa}} \left(1- e^{-\tau_{\nu}^{\rm sa}} \right) \ ,
\end{equation}
 
where, $j_{\nu}$ is the synchrotron emissivity,
$\alpha_{\nu}^{\rm sa}$ and $\tau_{\nu}^{\rm sa}$ are the self absorption (sa) coefficient and optical depth, respectively. For detail calculation of all these ingredients, the reader is referred to ~\cite{1979rpa..book.....R,1970ranp.book.....P}. 

\begin{figure}
    \centering
    \includegraphics[width=0.9\textwidth]{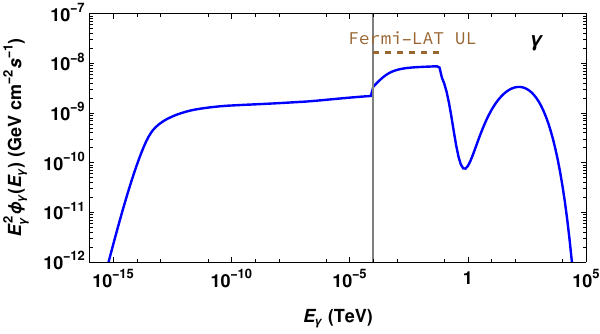}
    \caption{Multi-frequency synchrotron flux (left side of the gray line) together with gamma-ray flux from $\pi^0$ decay (right side of the gray line). The brown dashed line represents the Fermi-LAT  upper limit. }
    \label{fig:full_spectra}
\end{figure}

Apart from synchrotron, secondary electrons can also suffer from energy losses due to inverse Compton, bremsstrahlung and Coulomb losses. Among these, bremsstrahlung and Coulomb losses are found to be highly suppressed in dense CSM environment~\cite{Murase:2013kda}. However, the inverse Compton effect  depends on the choice of parameters (e.g., the density of target photons) and may become comparable to the synchrotron cooling~\cite[see][for details]{Murase:2013kda}.  For our benchmark scenario (Table~\ref{tab:parameters}), the cooling due to synchrotron loss is found to be dominant due to the choice of large value of $\epsilon_{\rm B}$. In such a simplistic scenario where the dominant process of energy losses is the same as the process responsible for radiation, i.e., synchrotron, one can  estimate the radiation spectrum using calorimetric approximation. The synchrotron flux (in units of $\rm TeV~cm^{-2}~s^{-1}$) for frequencies  above the characteristic frequency of secondary electrons ($\nu_{\rm c}$), i.e., for $\nu>\nu_{\rm c}$ is given by~\cite{Murase:2013kda}, 
\begin{equation}
  E_{\gamma}^2 \phi_{\gamma} =  \frac{1}{4 \pi d_{\rm L}^2} \frac{1}{12} \times \min[1, t_{\rm ad}/t_{\rm pp} ] \times E_{\rm p}^2  \frac{d \mathcal{N}_{\rm p}}{d E_{\rm p} d t} ~. 
\end{equation}

The characteristic frequency of secondary electrons ($\nu_{\rm c}$) depends on the parameters $\epsilon_{\rm B}$, $\dot{M}_{\rm W}$, $v_{\rm W}$,  and $v_{\rm sh}$ and found to be around $10$~THz for the benchmark case.

Fig.~\ref{fig:full_spectra} shows the synchrotron flux (for $\nu>\nu_{\rm c}$)  together with the gamma-ray flux from $\pi^{0}$ decay as a function of energy in TeV for the benchmark case. The  vertical gray line at energy $\sim 10^{-4}$~TeV separates the two different processes, i.e, synchrotron on the left side and $\pi^0$ decay on the right side of this line. The figure also shows the gamma-ray limit (brown dashed line) from Fermi-LAT. The synchrotron flux is found to be smaller than the gamma-ray flux from $\pi^0$ decay and remains much below the Fermi-LAT limit. Particularly,  the gamma-ray flux above $10^{-4}$~TeV is dominated by the contribution from $\pi^0$ decay and has negligible contribution from synchrotron radiation. Thus, the upper limit on the gamma-ray flux from $\pi^0$ decay remains unaffected by synchrotron radiation. 

We have also compared our synchrotron flux with existing estimation in the literature. The measured hard X-ray luminosity in the $0.3-79$~keV band is reported to be around $2.5\times 10^{40}~\rm erg~s^{-1}$~\cite{Grefenstette:2023dka}, which is consistent with  our prediction in this energy band $\sim 10^{40}~\rm erg~s^{-1}$. This prediction is also found to be consistent with the low frequency observation~\cite{2023arXiv230609311B}. Note that, primary electrons accelerated in the interaction of ejecta and CSM can also produce synchrotron radiation. However, their contribution is much smaller than that of secondary electrons in the energy range of our interest (Fermi-LAT energy range), \cite[see e.g.,][]{Murase:2018okz} and hence ignored. 

}
\section{Conclusion} \label{sec:conclusion}

In this work, we have estimated the upper limits of the gamma-ray and high energy neutrino fluxes from SN 2023ixf. Gamma-rays and high energy neutrinos can be produced by the interaction of shock accelerated protons with CSM protons. Evidence of dense CSM in the close proximity of the progenitor of SN 2023 has been inferred from observations. The upper limits of the fluxes are obtained by optimising the characteristic parameters of the shock-CSM model. The  gamma-ray flux ($7\times 10^{-9}~\rm GeV~cm^{-2}~s^{-1}$) is found to be consistent with the Fermi-LAT upper limit, while  the  neutrino flux ($4\times 10^{-9}~\rm GeV~cm^{-2}~s^{-1}$) is found to be about $2$ orders of magnitude smaller than the IceCube's upper limit. Hence, our constraints on the secondary fluxes are much stronger than the experimental upper limits.

{\color{black}
Gamma-rays in the Fermi-LAT energy band can also be produced via synchrotron radiation of secondary electrons. We have computed the synchrotron flux and found that it has negligible effect on our estimated upper limit of the gamma-ray flux from $\pi^{0}$ decay.}
On the other hand, high energy neutrinos could also be produced through gamma-ray annihilation~\citep{2023arXiv230506375H}. However, this process requires keV energy target photons and hence, not important.  

Our analysis is based on a conventional model of proton acceleration at forward shock which gives a power law distribution with a spectral index $2$ \cite{Murase:2010cu,Petropoulou:2016zar}.  However, effect of non-resonant instability or turbulence may lead to deviation from  $2$~\cite{10.1111/j.1365-2966.2004.08097.x,Cardillo:2015zda,2002APh....16..429B,1999ApJ...511L..53M}. For example, gamma-ray observations of the young SN remnant, Cas A infers a spectral index smaller than $2$~\cite{2017MNRAS.472.2956A}. Such a smaller spectral index would result in lower flux at higher energies. Since our goal is to obtain the maximum flux, we stick to the conventional power law model.

The prediction of the gamma-ray and neutrino fluxes also depends on the CSM profile. The origin of dense CSM in SN 2023ixf is not well understood. Note that some studies have shown a possible asymmetric nature of this CSM~\cite{2023arXiv230607964S,2023arXiv230701268V}.   For asymmetric CSM, the flux prediction would depend on the alignment of the CSM with respect to the line of sight.   Observations of SN 2023ixf suggest that most of the CSM may not be along the line of sight~\cite{2023arXiv230701268V} and therefore, the secondary fluxes can be expected to be smaller. Thus, our symmetric CSM model which is based on line of sight observations gives an estimate of the maximum possible fluxes.

We have also analysed the possibility of probing gamma-ray and neutrino fluxes for similar nearby future events with upcoming detectors, CTA and IceCube-Gen2. These detectors are found to have great potential to discover such fluxes. Indeed, they will be able to put stringent constraints on the particle acceleration in such SN environment.

\section*{Acknowledgements}
The author would like to thank Sovan Chakraborty and Madhurima Chakraborty for their useful comments on the paper.

\bibliographystyle{JHEP}
\bibliography{biblio}

\providecommand{\href}[2]{#2}\begingroup\raggedright\begin{thebibliography}{10}

\bibitem{2023arXiv230604721J}
W.V.~{Jacobson-Galan}, L.~{Dessart}, R.~{Margutti}, R.~{Chornock},
  R.J.~{Foley}, C.D.~{Kilpatrick} et~al., \emph{{SN 2023ixf in Messier 101:
  Photo-ionization of Dense, Close-in Circumstellar Material in a Nearby Type
  II Supernova}}, \href{https://doi.org/10.48550/arXiv.2306.04721}{\emph{arXiv
  e-prints} (2023) arXiv:2306.04721}
  [\href{https://arxiv.org/abs/2306.04721}{{\ttfamily 2306.04721}}].

\bibitem{2023arXiv230610119B}
K.A.~{Bostroem}, J.~{Pearson}, M.~{Shrestha}, D.J.~{Sand}, S.~{Valenti},
  S.W.~{Jha} et~al., \emph{{Early Spectroscopy and Dense Circumstellar Medium
  Interaction in SN\raisebox{-0.5ex}\textasciitilde2023ixf}},
  \href{https://doi.org/10.48550/arXiv.2306.10119}{\emph{arXiv e-prints} (2023)
  arXiv:2306.10119} [\href{https://arxiv.org/abs/2306.10119}{{\ttfamily
  2306.10119}}].

\bibitem{2023arXiv230703165H}
D.~{Hiramatsu}, D.~{Tsuna}, E.~{Berger}, K.~{Itagaki}, J.A.~{Goldberg},
  S.~{Gomez} et~al., \emph{{From Discovery to the First Month of the Type II
  Supernova 2023ixf: High and Variable Mass Loss in the Final Year Before
  Explosion}}, \href{https://doi.org/10.48550/arXiv.2307.03165}{\emph{arXiv
  e-prints} (2023) arXiv:2307.03165}
  [\href{https://arxiv.org/abs/2307.03165}{{\ttfamily 2307.03165}}].

\bibitem{2023arXiv230701268V}
S.S.~{Vasylyev}, Y.~{Yang}, A.V.~{Filippenko}, K.~{Patra}, T.G.~{Brink},
  L.~{Wang} et~al., \emph{{Early-time Spectropolarimetry of the Asymmetric Type
  II Supernova SN 2023ixf}},
  \href{https://doi.org/10.48550/arXiv.2307.01268}{\emph{arXiv e-prints} (2023)
  arXiv:2307.01268} [\href{https://arxiv.org/abs/2307.01268}{{\ttfamily
  2307.01268}}].

\bibitem{Schure:2013kya}
K.M.~Schure and A.R.~Bell, \emph{{Cosmic ray acceleration in young supernova
  remnants}}, \href{https://doi.org/10.1093/mnras/stt1371}{\emph{Mon. Not. Roy.
  Astron. Soc.} {\bfseries 435} (2013) 1174}
  [\href{https://arxiv.org/abs/1307.6575}{{\ttfamily 1307.6575}}].

\bibitem{Bell:2013kq}
A.~Bell, K.~Schure, B.~Reville and G.~Giacinti, \emph{{Cosmic ray acceleration
  and escape from supernova remnants}},
  \href{https://doi.org/10.1093/mnras/stt179}{\emph{Mon. Not. Roy. Astron.
  Soc.} {\bfseries 431} (2013) 415}
  [\href{https://arxiv.org/abs/1301.7264}{{\ttfamily 1301.7264}}].

\bibitem{Cardillo:2015zda}
M.~Cardillo, E.~Amato and P.~Blasi, \emph{{On the cosmic ray spectrum from type
  II Supernovae expanding in their red giant presupernova wind}},
  \href{https://doi.org/10.1016/j.astropartphys.2015.03.002}{\emph{Astropart.
  Phys.} {\bfseries 69} (2015) 1}
  [\href{https://arxiv.org/abs/1503.03001}{{\ttfamily 1503.03001}}].

\bibitem{Cristofari:2022kqv}
P.~Cristofari, P.~Blasi and D.~Caprioli, \emph{{Microphysics of Diffusive Shock
  Acceleration: Impact on the Spectrum of Accelerated Particles}},
  \href{https://doi.org/10.3847/1538-4357/ac6182}{\emph{Astrophys. J.}
  {\bfseries 930} (2022) 28}
  [\href{https://arxiv.org/abs/2203.15624}{{\ttfamily 2203.15624}}].

\bibitem{Murase:2010cu}
K.~Murase, T.A.~Thompson, B.C.~Lacki and J.F.~Beacom, \emph{{New Class of
  High-Energy Transients from Crashes of Supernova Ejecta with Massive
  Circumstellar Material Shells}},
  \href{https://doi.org/10.1103/PhysRevD.84.043003}{\emph{Phys. Rev. D}
  {\bfseries 84} (2011) 043003}
  [\href{https://arxiv.org/abs/1012.2834}{{\ttfamily 1012.2834}}].

\bibitem{Petropoulou:2016zar}
M.~Petropoulou, A.~Kamble and L.~Sironi, \emph{{Radio synchrotron emission from
  secondary electrons in interaction-powered supernovae}},
  \href{https://doi.org/10.1093/mnras/stw920}{\emph{Mon. Not. Roy. Astron.
  Soc.} {\bfseries 460} (2016) 44}
  [\href{https://arxiv.org/abs/1603.00891}{{\ttfamily 1603.00891}}].

\bibitem{Murase:2013kda}
K.~Murase, T.A.~Thompson and E.O.~Ofek, \emph{{Probing Cosmic-Ray Ion
  Acceleration with Radio-Submm and Gamma-Ray Emission from Interaction-Powered
  Supernovae}}, \href{https://doi.org/10.1093/mnras/stu384}{\emph{Mon. Not.
  Roy. Astron. Soc.} {\bfseries 440} (2014) 2528}
  [\href{https://arxiv.org/abs/1311.6778}{{\ttfamily 1311.6778}}].

\bibitem{Murase:2014bfa}
K.~Murase, K.~Kashiyama, K.~Kiuchi and I.~Bartos, \emph{{Gamma-Ray and Hard
  X-Ray Emission from Pulsar-Aided Supernovae as a Probe of Particle
  Acceleration in Embryonic Pulsar Wind Nebulae}},
  \href{https://doi.org/10.1088/0004-637X/805/1/82}{\emph{Astrophys. J.}
  {\bfseries 805} (2015) 82} [\href{https://arxiv.org/abs/1411.0619}{{\ttfamily
  1411.0619}}].

\bibitem{Murase:2017pfe}
K.~Murase, \emph{{New Prospects for Detecting High-Energy Neutrinos from Nearby
  Supernovae}}, \href{https://doi.org/10.1103/PhysRevD.97.081301}{\emph{Phys.
  Rev. D} {\bfseries 97} (2018) 081301}
  [\href{https://arxiv.org/abs/1705.04750}{{\ttfamily 1705.04750}}].

\bibitem{Murase:2018okz}
K.~Murase, A.~Franckowiak, K.~Maeda, R.~Margutti and J.F.~Beacom,
  \emph{{High-Energy Emission from Interacting Supernovae: New Constraints on
  Cosmic-Ray Acceleration in Dense Circumstellar Environments}},
  \href{https://doi.org/10.3847/1538-4357/ab0422}{\emph{Astrophys. J.}
  {\bfseries 874} (2019) 80}
  [\href{https://arxiv.org/abs/1807.01460}{{\ttfamily 1807.01460}}].

\bibitem{Petropoulou:2017ymv}
M.~Petropoulou, S.~Coenders, G.~Vasilopoulos, A.~Kamble and L.~Sironi,
  \emph{{Point-source and diffuse high-energy neutrino emission from Type IIn
  supernovae}}, \href{https://doi.org/10.1093/mnras/stx1251}{\emph{Mon. Not.
  Roy. Astron. Soc.} {\bfseries 470} (2017) 1881}
  [\href{https://arxiv.org/abs/1705.06752}{{\ttfamily 1705.06752}}].

\bibitem{2023ATel16075....1M}
G.~{Marti-Devesa}, \emph{{Fermi-LAT gamma-ray observations of SN 2023ixf}},
  {\emph{The Astronomer's Telegram} {\bfseries 16075} (2023) 1}.

\bibitem{2023ATel16043....1T}
J.~{Thwaites}, J.~{Vandenbroucke}, M.~{Santander} and {IceCube Collaboration},
  \emph{{SN 2023ixf: Upper limits from a neutrino search with IceCube}},
  {\emph{The Astronomer's Telegram} {\bfseries 16043} (2023) 1}.

\bibitem{Smith:2014txa}
N.~Smith, \emph{{Mass Loss: Its Effect on the Evolution and Fate of High-Mass
  Stars}},
  \href{https://doi.org/10.1146/annurev-astro-081913-040025}{\emph{Ann. Rev.
  Astron. Astrophys.} {\bfseries 52} (2014) 487}
  [\href{https://arxiv.org/abs/1402.1237}{{\ttfamily 1402.1237}}].

\bibitem{2013MNRAS.430.1801M}
J.C.~{Mauerhan}, N.~{Smith}, A.V.~{Filippenko}, K.B.~{Blanchard},
  P.K.~{Blanchard}, C.F.E.~{Casper} et~al., \emph{{The unprecedented 2012
  outburst of SN 2009ip: a luminous blue variable star becomes a true
  supernova}}, \href{https://doi.org/10.1093/mnras/stt009}{\emph{mnras}
  {\bfseries 430} (2013) 1801}
  [\href{https://arxiv.org/abs/1209.6320}{{\ttfamily 1209.6320}}].

\bibitem{Morozova:2016efp}
V.~Morozova, A.L.~Piro and S.~Valenti, \emph{{Unifying Type II Supernova Light
  Curves with Dense Circumstellar Material}},
  \href{https://doi.org/10.3847/1538-4357/aa6251}{\emph{Astrophys. J.}
  {\bfseries 838} (2017) 28}
  [\href{https://arxiv.org/abs/1610.08054}{{\ttfamily 1610.08054}}].

\bibitem{Yaron:2017umb}
O.~Yaron et~al., \emph{{Confined Dense Circumstellar Material Surrounding a
  Regular Type II Supernova: The Unique Flash-Spectroscopy Event of SN
  2013fs}}, \href{https://doi.org/10.1038/nphys4025}{\emph{Nature Phys.}
  {\bfseries 13} (2017) 510}
  [\href{https://arxiv.org/abs/1701.02596}{{\ttfamily 1701.02596}}].

\bibitem{Owocki_2004}
S.P.~Owocki, K.G.~Gayley and N.J.~Shaviv, \emph{A porosity-length formalism for
  photon-tiring-limited mass loss from stars above the eddington limit},
  \href{https://doi.org/10.1086/424910}{\emph{The Astrophysical Journal}
  {\bfseries 616} (2004) 525}.

\bibitem{10.1093/mnras/stx2251}
S.P.~Owocki, R.H.D.~Townsend and E.~Quataert, \emph{{Super-Eddington stellar
  winds: unifying radiative-enthalpy versus flux-driven models}},
  \href{https://doi.org/10.1093/mnras/stx2251}{\emph{Monthly Notices of the
  Royal Astronomical Society} {\bfseries 472} (2017) 3749}
  [\href{https://arxiv.org/abs/https://academic.oup.com/mnras/article-pdf/472/3/3749/20331271/stx2251.pdf}{{\ttfamily
  https://academic.oup.com/mnras/article-pdf/472/3/3749/20331271/stx2251.pdf}}].

\bibitem{10.1093/mnras/stw365}
E.~Quataert, R.~Fernández, D.~Kasen, H.~Klion and B.~Paxton,
  \emph{{Super-Eddington stellar winds driven by near-surface energy
  deposition}}, \href{https://doi.org/10.1093/mnras/stw365}{\emph{Monthly
  Notices of the Royal Astronomical Society} {\bfseries 458} (2016) 1214}
  [\href{https://arxiv.org/abs/https://academic.oup.com/mnras/article-pdf/458/2/1214/18239162/stw365.pdf}{{\ttfamily
  https://academic.oup.com/mnras/article-pdf/458/2/1214/18239162/stw365.pdf}}].

\bibitem{10.1093/mnras/stx1314}
J.~Fuller, \emph{{Pre-supernova outbursts via wave heating in massive stars –
  I. Red supergiants}},
  \href{https://doi.org/10.1093/mnras/stx1314}{\emph{Monthly Notices of the
  Royal Astronomical Society} {\bfseries 470} (2017) 1642}
  [\href{https://arxiv.org/abs/https://academic.oup.com/mnras/article-pdf/470/2/1642/18023178/stx1314.pdf}{{\ttfamily
  https://academic.oup.com/mnras/article-pdf/470/2/1642/18023178/stx1314.pdf}}].

\bibitem{2021ApJ...906....3W}
S.~{Wu} and J.~{Fuller}, \emph{{A Diversity of Wave-driven Presupernova
  Outbursts}}, \href{https://doi.org/10.3847/1538-4357/abc87c}{\emph{apj}
  {\bfseries 906} (2021) 3} [\href{https://arxiv.org/abs/2011.05453}{{\ttfamily
  2011.05453}}].

\bibitem{10.1093/mnras/staa255}
E.R.~Beasor, B.~Davies, N.~Smith, J.T.~van Loon, R.D.~Gehrz and D.F.~Figer,
  \emph{{A new mass-loss rate prescription for red supergiants}},
  \href{https://doi.org/10.1093/mnras/staa255}{\emph{Monthly Notices of the
  Royal Astronomical Society} {\bfseries 492} (2020) 5994}
  [\href{https://arxiv.org/abs/https://academic.oup.com/mnras/article-pdf/492/4/5994/32445077/staa255.pdf}{{\ttfamily
  https://academic.oup.com/mnras/article-pdf/492/4/5994/32445077/staa255.pdf}}].

\bibitem{2023arXiv230700727M}
T.~{Matsuoka} and R.~{Sawada}, \emph{{Binary Interaction Can Yield a Diversity
  of Circumstellar Media around Type II Supernova Progenitors}},
  \href{https://doi.org/10.48550/arXiv.2307.00727}{\emph{arXiv e-prints} (2023)
  arXiv:2307.00727} [\href{https://arxiv.org/abs/2307.00727}{{\ttfamily
  2307.00727}}].

\bibitem{2023arXiv230615270S}
N.~{Soker}, \emph{{A pre-explosion effervescent zone for the circumstellar
  material in SN 2023ixf}},
  \href{https://doi.org/10.48550/arXiv.2306.15270}{\emph{arXiv e-prints} (2023)
  arXiv:2306.15270} [\href{https://arxiv.org/abs/2306.15270}{{\ttfamily
  2306.15270}}].

\bibitem{2023arXiv230607964S}
N.~{Smith}, J.~{Pearson}, D.J.~{Sand}, I.~{Ilyin}, K.A.~{Bostroem},
  G.~{Hosseinzadeh} et~al., \emph{{High resolution spectroscopy of
  SN\raisebox{-0.5ex}\textasciitilde2023ixf's first week: Engulfing the
  Asymmetric Circumstellar Material}},
  \href{https://doi.org/10.48550/arXiv.2306.07964}{\emph{arXiv e-prints} (2023)
  arXiv:2306.07964} [\href{https://arxiv.org/abs/2306.07964}{{\ttfamily
  2306.07964}}].

\bibitem{2023arXiv230608678J}
J.E.~{Jencson}, J.~{Pearson}, E.R.~{Beasor}, R.M.~{Lau}, J.E.~{Andrews},
  K.A.~{Bostroem} et~al., \emph{{A Luminous Red Supergiant and Dusty
  Long-period Variable Progenitor for SN 2023ixf}},
  \href{https://doi.org/10.48550/arXiv.2306.08678}{\emph{arXiv e-prints} (2023)
  arXiv:2306.08678} [\href{https://arxiv.org/abs/2306.08678}{{\ttfamily
  2306.08678}}].

\bibitem{2023arXiv230514447P}
J.L.~{Pledger} and M.M.~{Shara}, \emph{{Possible detection of the progenitor of
  the Type II supernova SN2023ixf}},
  \href{https://doi.org/10.48550/arXiv.2305.14447}{\emph{arXiv e-prints} (2023)
  arXiv:2305.14447} [\href{https://arxiv.org/abs/2305.14447}{{\ttfamily
  2305.14447}}].

\bibitem{Mauerhan:2018wes}
J.C.~Mauerhan, W.~Zheng, T.~Brink, M.L.~Graham, I.~Shivvers, K.~Clubb et~al.,
  \emph{{Stripped-envelope supernova SN 2004dk is now interacting with
  hydrogen-rich circumstellar material}},
  \href{https://doi.org/10.1093/mnras/sty1307}{\emph{Mon. Not. Roy. Astron.
  Soc.} {\bfseries 478} (2018) 5050}
  [\href{https://arxiv.org/abs/1803.07051}{{\ttfamily 1803.07051}}].

\bibitem{2022arXiv221203313B}
R.J.~{Bruch}, A.~{Gal-Yam}, O.~{Yaron}, P.~{Chen}, N.L.~{Strotjohann},
  I.~{Irani} et~al., \emph{{The prevalence and influence of circumstellar
  material around hydrogen-rich supernova progenitors}},
  \href{https://doi.org/10.48550/arXiv.2212.03313}{\emph{arXiv e-prints} (2022)
  arXiv:2212.03313} [\href{https://arxiv.org/abs/2212.03313}{{\ttfamily
  2212.03313}}].

\bibitem{Milisavljevic:2015bli}
D.~Milisavljevic et~al., \emph{{Metamorphosis of SN 2014C: Delayed Interaction
  Between a Hydrogen Poor Core-collapse Supernova and a Nearby Circumstellar
  Shell}}, \href{https://doi.org/10.1088/0004-637X/815/2/120}{\emph{Astrophys.
  J.} {\bfseries 815} (2015) 120}
  [\href{https://arxiv.org/abs/1511.01907}{{\ttfamily 1511.01907}}].

\bibitem{2022JCAP...08..011S}
P.~{Sarmah}, S.~{Chakraborty}, I.~{Tamborra} and K.~{Auchettl}, \emph{{High
  energy particles from young supernovae: gamma-ray and neutrino connections}},
  \href{https://doi.org/10.1088/1475-7516/2022/08/011}{\emph{jcap} {\bfseries
  2022} (2022) 011} [\href{https://arxiv.org/abs/2204.03663}{{\ttfamily
  2204.03663}}].

\bibitem{2023arXiv230313576S}
P.~{Sarmah}, S.~{Chakraborty}, I.~{Tamborra} and K.~{Auchettl},
  \emph{{Gamma-rays and neutrinos from supernovae of Type Ib/c with late time
  emission}}, \href{https://doi.org/10.48550/arXiv.2303.13576}{\emph{arXiv
  e-prints} (2023) arXiv:2303.13576}
  [\href{https://arxiv.org/abs/2303.13576}{{\ttfamily 2303.13576}}].

\bibitem{IceCube:2020wum}
{\scshape IceCube} collaboration, \emph{{The IceCube high-energy starting event
  sample: Description and flux characterization with 7.5 years of data}},
  \href{https://arxiv.org/abs/2011.03545}{{\ttfamily 2011.03545}}.

\bibitem{Ackermann:2014usa}
{\scshape Fermi-LAT} collaboration, \emph{{The spectrum of isotropic diffuse
  gamma-ray emission between 100 MeV and 820 GeV}},
  \href{https://doi.org/10.1088/0004-637X/799/1/86}{\emph{Astrophys. J.}
  {\bfseries 799} (2015) 86} [\href{https://arxiv.org/abs/1410.3696}{{\ttfamily
  1410.3696}}].

\bibitem{Pitik:2021dyf}
T.~Pitik, I.~Tamborra, C.R.~Angus and K.~Auchettl, \emph{{Is the High-energy
  Neutrino Event IceCube-200530A Associated with a Hydrogen-rich Superluminous
  Supernova?}},
  \href{https://doi.org/10.3847/1538-4357/ac5ab1}{\emph{Astrophys. J.}
  {\bfseries 929} (2022) 163}
  [\href{https://arxiv.org/abs/2110.06944}{{\ttfamily 2110.06944}}].

\bibitem{Pitik:2023vcg}
T.~Pitik, I.~Tamborra, M.~Lincetto and A.~Franckowiak, \emph{{Optically
  Informed Searches of High-Energy Neutrinos from Interaction-Powered
  Supernovae}},  \href{https://arxiv.org/abs/2306.01833}{{\ttfamily
  2306.01833}}.

\bibitem{Zirakashvili:2015tda}
V.N.~Zarikashvili and V.~Ptuskin, \emph{{Type IIn supernovae as sources of high
  energy neutrinos}}, \href{https://doi.org/10.22323/1.236.0472}{\emph{PoS}
  {\bfseries ICRC2015} (2016) 472}
  [\href{https://arxiv.org/abs/1505.08144}{{\ttfamily 1505.08144}}].

\bibitem{2018MNRAS.479.4470M}
A.~{Marcowith}, V.V.~{Dwarkadas}, M.~{Renaud}, V.~{Tatischeff} and
  G.~{Giacinti}, \emph{{Core-collapse supernovae as cosmic ray sources}},
  \href{https://doi.org/10.1093/mnras/sty1743}{\emph{mnras} {\bfseries 479}
  (2018) 4470} [\href{https://arxiv.org/abs/1806.09700}{{\ttfamily
  1806.09700}}].

\bibitem{2014NuPhS.256...94M}
A.~{Marcowith}, M.~{Renaud}, V.~{Dwarkadas} and V.~{Tatischeff},
  \emph{{Cosmic-ray acceleration and gamma-ray signals from radio
  supernov{\ae}}},
  \href{https://doi.org/10.1016/j.nuclphysbps.2014.10.011}{\emph{Nuclear
  Physics B Proceedings Supplements} {\bfseries 256} (2014) 94}
  [\href{https://arxiv.org/abs/1409.3670}{{\ttfamily 1409.3670}}].

\bibitem{2022MNRAS.511.3321C}
P.~{Cristofari}, A.~{Marcowith}, M.~{Renaud}, V.V.~{Dwarkadas},
  V.~{Tatischeff}, G.~{Giacinti} et~al., \emph{{The first days of Type II-P
  core collapse supernovae in the gamma-ray range}},
  \href{https://doi.org/10.1093/mnras/stac217}{\emph{mnras} {\bfseries 511}
  (2022) 3321} [\href{https://arxiv.org/abs/2201.09583}{{\ttfamily
  2201.09583}}].

\bibitem{2020MNRAS.494.2760C}
P.~{Cristofari}, M.~{Renaud}, A.~{Marcowith}, V.V.~{Dwarkadas} and
  V.~{Tatischeff}, \emph{{Time-dependent high-energy gamma-ray signal from
  accelerated particles in core-collapse supernovae: the case of SN 1993J}},
  \href{https://doi.org/10.1093/mnras/staa984}{\emph{mnras} {\bfseries 494}
  (2020) 2760} [\href{https://arxiv.org/abs/2004.02650}{{\ttfamily
  2004.02650}}].

\bibitem{2023ApJ...955L...9G}
D.~{Guetta}, A.~{Langella}, S.~{Gagliardini} and M.D.~{Valle}, \emph{{Low- and
  High-energy Neutrinos from SN 2023ixf in M101}},
  \href{https://doi.org/10.3847/2041-8213/acf573}{\emph{apjl} {\bfseries 955}
  (2023) L9} [\href{https://arxiv.org/abs/2306.14717}{{\ttfamily 2306.14717}}].

\bibitem{2011ExA....32..193A}
M.~{Actis}, G.~{Agnetta}, F.~{Aharonian}, A.~{Akhperjanian}, J.~{Aleksi{\'c}},
  E.~{Aliu} et~al., \emph{{Design concepts for the Cherenkov Telescope Array
  CTA: an advanced facility for ground-based high-energy gamma-ray astronomy}},
  \href{https://doi.org/10.1007/s10686-011-9247-0}{\emph{Experimental
  Astronomy} {\bfseries 32} (2011) 193}
  [\href{https://arxiv.org/abs/1008.3703}{{\ttfamily 1008.3703}}].

\bibitem{IceCube-Gen2:2020qha}
{\scshape IceCube-Gen2} collaboration, \emph{{IceCube-Gen2: the window to the
  extreme Universe}}, \href{https://doi.org/10.1088/1361-6471/abbd48}{\emph{J.
  Phys. G} {\bfseries 48} (2021) 060501}
  [\href{https://arxiv.org/abs/2008.04323}{{\ttfamily 2008.04323}}].

\bibitem{Kelner:2006tc}
S.R.~Kelner, F.A.~Aharonian and V.V.~Bugayov, \emph{{Energy spectra of
  gamma-rays, electrons and neutrinos produced at proton-proton interactions in
  the very high energy regime}},
  \href{https://doi.org/10.1103/PhysRevD.74.034018}{\emph{Phys. Rev. D}
  {\bfseries 74} (2006) 034018}
  [\href{https://arxiv.org/abs/astro-ph/0606058}{{\ttfamily
  astro-ph/0606058}}].

\bibitem{10.1111/j.1365-2966.2004.08097.x}
A.R.~Bell, \emph{{Turbulent amplification of magnetic field and diffusive shock
  acceleration of cosmic rays}},
  \href{https://doi.org/10.1111/j.1365-2966.2004.08097.x}{\emph{Monthly Notices
  of the Royal Astronomical Society} {\bfseries 353} (2004) 550}
  [\href{https://arxiv.org/abs/https://academic.oup.com/mnras/article-pdf/353/2/550/3869038/353-2-550.pdf}{{\ttfamily
  https://academic.oup.com/mnras/article-pdf/353/2/550/3869038/353-2-550.pdf}}].

\bibitem{2009MNRAS.392.1591A}
E.~{Amato} and P.~{Blasi}, \emph{{A kinetic approach to cosmic-ray-induced
  streaming instability at supernova shocks}},
  \href{https://doi.org/10.1111/j.1365-2966.2008.14200.x}{\emph{mnras}
  {\bfseries 392} (2009) 1591}
  [\href{https://arxiv.org/abs/0806.1223}{{\ttfamily 0806.1223}}].

\bibitem{10.1093/mnras/stt179}
A.R.~Bell, K.M.~Schure, B.~Reville and G.~Giacinti, \emph{{Cosmic-ray
  acceleration and escape from supernova remnants}},
  \href{https://doi.org/10.1093/mnras/stt179}{\emph{Monthly Notices of the
  Royal Astronomical Society} {\bfseries 431} (2013) 415}
  [\href{https://arxiv.org/abs/https://academic.oup.com/mnras/article-pdf/431/1/415/18243010/stt179.pdf}{{\ttfamily
  https://academic.oup.com/mnras/article-pdf/431/1/415/18243010/stt179.pdf}}].

\bibitem{Schure:2012du}
K.M.~Schure, A.R.~Bell, L.O.~Drury and A.M.~Bykov, \emph{{Diffusive shock
  acceleration and magnetic field amplification}},
  \href{https://doi.org/10.1007/s11214-012-9871-7}{\emph{Space Sci. Rev.}
  {\bfseries 173} (2012) 491}
  [\href{https://arxiv.org/abs/1203.1637}{{\ttfamily 1203.1637}}].

\bibitem{2009ApJ...695..825I}
T.~{Inoue}, R.~{Yamazaki} and S.-i.~{Inutsuka}, \emph{{Turbulence and Magnetic
  Field Amplification in Supernova Remnants: Interactions Between a Strong
  Shock Wave and Multiphase Interstellar Medium}},
  \href{https://doi.org/10.1088/0004-637X/695/2/825}{\emph{apj} {\bfseries 695}
  (2009) 825} [\href{https://arxiv.org/abs/0901.0486}{{\ttfamily 0901.0486}}].

\bibitem{2012ApJ...744...71I}
T.~{Inoue}, R.~{Yamazaki}, S.-i.~{Inutsuka} and Y.~{Fukui}, \emph{{Toward
  Understanding the Cosmic-Ray Acceleration at Young Supernova Remnants
  Interacting with Interstellar Clouds: Possible Applications to RX
  J1713.7-3946}}, \href{https://doi.org/10.1088/0004-637X/744/1/71}{\emph{apj}
  {\bfseries 744} (2012) 71} [\href{https://arxiv.org/abs/1106.3080}{{\ttfamily
  1106.3080}}].

\bibitem{2023arXiv230609311B}
E.~{Berger}, G.K.~{Keating}, R.~{Margutti}, K.~{Maeda}, K.D.~{Alexander},
  Y.~{Cendes} et~al., \emph{{Millimeter Observations of the Type II SN2023ixf:
  Constraints on the Proximate Circumstellar Medium}},
  \href{https://doi.org/10.48550/arXiv.2306.09311}{\emph{arXiv e-prints} (2023)
  arXiv:2306.09311} [\href{https://arxiv.org/abs/2306.09311}{{\ttfamily
  2306.09311}}].

\bibitem{2014ApJ...781...42O}
E.O.~{Ofek}, A.~{Zoglauer}, S.E.~{Boggs}, N.M.~{Barri{\'e}re}, S.P.~{Reynolds},
  C.L.~{Fryer} et~al., \emph{{SN 2010jl: Optical to Hard X-Ray Observations
  Reveal an Explosion Embedded in a Ten Solar Mass Cocoon}},
  \href{https://doi.org/10.1088/0004-637X/781/1/42}{\emph{apj} {\bfseries 781}
  (2014) 42} [\href{https://arxiv.org/abs/1307.2247}{{\ttfamily 1307.2247}}].

\bibitem{2009ApJ...690.1839C}
P.~{Chandra}, C.J.~{Stockdale}, R.A.~{Chevalier}, S.D.~{Van Dyk}, A.~{Ray},
  M.T.~{Kelley} et~al., \emph{{Eleven Years of Radio Monitoring of the type IIn
  Supernova SN 1995N}},
  \href{https://doi.org/10.1088/0004-637X/690/2/1839}{\emph{apj} {\bfseries
  690} (2009) 1839} [\href{https://arxiv.org/abs/0809.2810}{{\ttfamily
  0809.2810}}].

\bibitem{2013ApJ...768...47O}
E.O.~{Ofek}, L.~{Lin}, C.~{Kouveliotou}, G.~{Younes},
  E.~{G{\"o}{\v{g}}{\"u}{\c{s}}}, M.M.~{Kasliwal} et~al., \emph{{SN 2009ip:
  Constraints on the Progenitor Mass-loss Rate}},
  \href{https://doi.org/10.1088/0004-637X/768/1/47}{\emph{apj} {\bfseries 768}
  (2013) 47} [\href{https://arxiv.org/abs/1303.3894}{{\ttfamily 1303.3894}}].

\bibitem{2000ApJ...536..195C}
R.A.~{Chevalier} and Z.-Y.~{Li}, \emph{{Wind Interaction Models for Gamma-Ray
  Burst Afterglows: The Case for Two Types of Progenitors}},
  \href{https://doi.org/10.1086/308914}{\emph{apj} {\bfseries 536} (2000) 195}
  [\href{https://arxiv.org/abs/astro-ph/9908272}{{\ttfamily
  astro-ph/9908272}}].

\bibitem{Caprioli:2008aq}
D.~Caprioli, P.~Blasi, E.~Amato and M.~Vietri, \emph{{Dynamical effects of
  self-generated magnetic fields in cosmic ray modified shocks}},
  \href{https://doi.org/10.1086/589505}{\emph{Astrophys. J. Lett.} {\bfseries
  679} (2008) L139} [\href{https://arxiv.org/abs/0804.2884}{{\ttfamily
  0804.2884}}].

\bibitem{Sturner_1997}
S.J.~Sturner, J.G.~Skibo, C.D.~Dermer and J.R.~Mattox, \emph{Temporal evolution
  of nonthermal spectra from supernova remnants},
  \href{https://doi.org/10.1086/304894}{\emph{The Astrophysical Journal}
  {\bfseries 490} (1997) 619}.

\bibitem{1997ApJ...490..619S}
S.J.~{Sturner}, J.G.~{Skibo}, C.D.~{Dermer} and J.R.~{Mattox}, \emph{{Temporal
  Evolution of Nonthermal Spectra from Supernova Remnants}},
  \href{https://doi.org/10.1086/304894}{\emph{apj} {\bfseries 490} (1997) 619}.

\bibitem{1995A&A...295..613M}
A.~{Mastichiadis} and J.G.~{Kirk}, \emph{{Self-consistent particle acceleration
  in active galactic nuclei.}}, {\emph{aap} {\bfseries 295} (1995) 613}.

\bibitem{PhysRevD.90.023010}
M.~Ahlers and K.~Murase, \emph{Probing the galactic origin of the icecube
  excess with gamma rays},
  \href{https://doi.org/10.1103/PhysRevD.90.023010}{\emph{Phys. Rev. D}
  {\bfseries 90} (2014) 023010}.

\bibitem{Chevalier:2006vy}
R.A.~Chevalier and C.~Fransson, \emph{{Cicumstellar Emission from Type Ib and
  Ic Supernovae}}, \href{https://doi.org/10.1086/507606}{\emph{Astrophys. J.}
  {\bfseries 651} (2006) 381}
  [\href{https://arxiv.org/abs/astro-ph/0607196}{{\ttfamily
  astro-ph/0607196}}].

\bibitem{2010ApJ...725..922S}
A.M.~{Soderberg}, A.~{Brunthaler}, E.~{Nakar}, R.A.~{Chevalier} and
  M.F.~{Bietenholz}, \emph{{Radio and X-ray Observations of the Type Ic SN
  2007gr Reveal an Ordinary, Non-relativistic Explosion}},
  \href{https://doi.org/10.1088/0004-637X/725/1/922}{\emph{apj} {\bfseries 725}
  (2010) 922} [\href{https://arxiv.org/abs/1005.1932}{{\ttfamily 1005.1932}}].

\bibitem{2016ApJ...818..111K}
A.~{Kamble}, R.~{Margutti}, A.M.~{Soderberg}, S.~{Chakraborti}, C.~{Fransson},
  R.~{Chevalier} et~al., \emph{{Progenitors of Type IIB Supernovae in the Light
  of Radio and X-Rays from SN 2013DF}},
  \href{https://doi.org/10.3847/0004-637X/818/2/111}{\emph{apj} {\bfseries 818}
  (2016) 111} [\href{https://arxiv.org/abs/1504.07988}{{\ttfamily
  1504.07988}}].

\bibitem{2017MNRAS.472.2956A}
M.L.~{Ahnen}, S.~{Ansoldi}, L.A.~{Antonelli}, C.~{Arcaro}, A.~{Babi{\'c}},
  B.~{Banerjee} et~al., \emph{{A cut-off in the TeV gamma-ray spectrum of the
  SNR Cassiopeia A}}, \href{https://doi.org/10.1093/mnras/stx2079}{\emph{mnras}
  {\bfseries 472} (2017) 2956}
  [\href{https://arxiv.org/abs/1707.01583}{{\ttfamily 1707.01583}}].

\bibitem{1979rpa..book.....R}
G.B.~{Rybicki} and A.P.~{Lightman}, \emph{{Radiative processes in
  astrophysics}} (1979).

\bibitem{1970ranp.book.....P}
A.G.~{Pacholczyk}, \emph{{Radio astrophysics. Nonthermal processes in galactic
  and extragalactic sources}} (1970).

\bibitem{Grefenstette:2023dka}
B.W.~Grefenstette, M.~Brightman, H.P.~Earnshaw, F.A.~Harrison and R.~Margutti,
  \emph{{Early Hard X-Rays from the Nearby Core-collapse Supernova SN
  2023ixf}}, \href{https://doi.org/10.3847/2041-8213/acdf4e}{\emph{Astrophys.
  J. Lett.} {\bfseries 952} (2023) L3}
  [\href{https://arxiv.org/abs/2306.04827}{{\ttfamily 2306.04827}}].

\bibitem{2023arXiv230506375H}
D.~{Hooper} and K.~{Plant}, \emph{{A Leptonic Model for Neutrino Emission From
  Active Galactic Nuclei}},
  \href{https://doi.org/10.48550/arXiv.2305.06375}{\emph{arXiv e-prints} (2023)
  arXiv:2305.06375} [\href{https://arxiv.org/abs/2305.06375}{{\ttfamily
  2305.06375}}].

\bibitem{2002APh....16..429B}
P.~{Blasi}, \emph{{A semi-analytical approach to non-linear shock
  acceleration}},
  \href{https://doi.org/10.1016/S0927-6505(01)00127-X}{\emph{Astroparticle
  Physics} {\bfseries 16} (2002) 429}
  [\href{https://arxiv.org/abs/astro-ph/0104064}{{\ttfamily
  astro-ph/0104064}}].

\bibitem{1999ApJ...511L..53M}
M.A.~{Malkov}, \emph{{Asymptotic Particle Spectra and Plasma Flows at Strong
  Shocks}}, \href{https://doi.org/10.1086/311825}{\emph{apjl} {\bfseries 511}
  (1999) L53} [\href{https://arxiv.org/abs/astro-ph/9807097}{{\ttfamily
  astro-ph/9807097}}].

\end{thebibliography}\endgroup

\end{document}